\documentclass[aps,prb,reprint,amsmath,amssymb,superscriptaddress,showpacs]{revtex4-2}
\usepackage{graphicx}
\usepackage{dcolumn}
\usepackage{bm}
\usepackage[colorlinks,linkcolor=blue,anchorcolor=blue,citecolor=blue,urlcolor=blue,filecolor=blue,menucolor=blue,runcolor=blue]{hyperref}

\begin{document}
\title{Large ferromagnetic-like band splitting in ultrathin SmC$_{6}$ films}
\author{Hao Zheng}
\affiliation{School of Physics, Zhejiang University, Hangzhou 310058, China}
\affiliation{Center for Correlated Matter, Zhejiang University, Hangzhou 310058, China}
\author{Yifu Xu}
\affiliation{School of Physics, Zhejiang University, Hangzhou 310058, China}
\affiliation{Center for Correlated Matter, Zhejiang University, Hangzhou 310058, China}
\author{Guowei Yang}
\affiliation{School of Physics, Zhejiang University, Hangzhou 310058, China}
\affiliation{Center for Correlated Matter, Zhejiang University, Hangzhou 310058, China}
\author{Ze Pan}
\affiliation{School of Physics, Zhejiang University, Hangzhou 310058, China}
\affiliation{Center for Correlated Matter, Zhejiang University, Hangzhou 310058, China}
\author{Yi Wu}
\affiliation{School of Physics, Zhejiang University, Hangzhou 310058, China}
\affiliation{Center for Correlated Matter, Zhejiang University, Hangzhou 310058, China}
\author{Yuan Zheng}
\affiliation{School of Physics, Zhejiang University, Hangzhou 310058, China}
\author{Tulai Sun}
\email {tlsun2020@zjut.edu.cn}
\affiliation{Center for Electron Microscopy, Zhejiang University of Technology, Hanghzhou 310014, China}
\author{Jiefeng Cao}
\affiliation{Shanghai Synchrotron Radiation Facility, Shanghai Advanced Research Institute, Chinese Academy of Sciences, Shanghai 201204, China}
\author{Yi-feng Yang}
\affiliation{Beijing National Laboratory for Condensed Matter Physics, Institute of Physics, Chinese Academy of Science, Beijing 100190, China}
\author{Ming Shi}
\affiliation{School of Physics, Zhejiang University, Hangzhou 310058, China}
\affiliation{Center for Correlated Matter, Zhejiang University, Hangzhou 310058, China}
\author{Chao Cao}
\email {ccao@zju.edu.cn}
\affiliation{School of Physics, Zhejiang University, Hangzhou 310058, China}
\affiliation{Center for Correlated Matter, Zhejiang University, Hangzhou 310058, China}
\author{Yang Liu}
\email {yangliuphys@zju.edu.cn}
\affiliation{School of Physics, Zhejiang University, Hangzhou 310058, China}
\affiliation{Center for Correlated Matter, Zhejiang University, Hangzhou 310058, China}
\affiliation{Collaborative Innovation Center of Advanced Microstructures, Nanjing University, Nanjing 210093, China}
\date{\today}%
\addcontentsline{toc}{chapter}{Abstract}

\begin{abstract}
Two-dimensional (2D) magnetic materials provide a unique platform for exploring quantum phases from magnetic order in reduced dimensions. While there have been extensive studies on 2D magnetic materials based on $3d$ electrons, experimental studies on $4f$-electron counterparts are far fewer, particularly on their electronic structure.  
In this study, we report the successful synthesis of ultrathin SmC$_{6}$ films using molecular beam epitaxy. 
Utilizing $in$ $situ$ angle-resolved photoemission spectroscopy (ARPES), we uncover a large band splitting in the valence bands, which we attribute to the ferromagnetic order driven by exchange couplings between Sm $4f$ moments and conduction electrons. 
Despite the small magnetic moment of Sm, the observed splitting is comparable to those of Eu- and Gd-based systems with much larger local moments. Interestingly, the surface state also exhibits splitting with similar magnitude and can be eliminated by overannealing, while the valence bands with ferromagnetic-like splittings remain robust. Our work provides spectroscopic insight to understand the electronic origin of magnetic order in Sm-based compounds. Our study also offers a platform to study 2D magnetic materials based on $4f$ electrons. 
\end{abstract}

\maketitle

\section{Introduction}
The discovery of two-dimensional (2D) magnetic materials has opened avenues for exploring magnetism and functionalities in reduced dimensions \cite{2Dreview2018,2Dreview2019}. Examples such as Cr$_{2}$Ge$_{2}$Te$_{6}$, CrI$_{3}$ and VSe$_{2}$ \cite{Cr2Ge2Te6nature2017,CrI3nature2017,VSe2NN2018} have demonstrated that 2D systems can sustain long-range magnetic order, enabled by magnetic anisotropy in 2D. Unlike their three-dimensional counterparts, 2D magnets exhibit properties that are highly tunable through external parameters such as gating \cite{FGTgatingnature2018,CrI3gatingNN2018}, strain \cite{VX2strainACS2012,FGTstrainACS2020} and thickness \cite{CrI3nature2017,Cr2Te3NC2023,CrTe2NC2021}, enabling precise manipulation of their magnetic and electronic properties.
These characters make 2D magnetic materials exciting platforms for both the fundamental research and applications in spintronic devices. 
Rare-earth-based 2D magnetic materials represent a particularly interesting class of systems, due to the large local moments and strong spin-orbit coupling from $4f$ electrons, tunable interactions between the $4f$ and conduction electrons, as well as significant magnetocrystalline anisotropy \cite{2Dreview2021,RESi2-NC2018,CeSiI-nature2024}. These features provide opportunites to explore the coupling between spin, orbital, and lattice degrees of freedom, which are essential for developing magnetic and electronic functionalities for 2D materials. However, compared to the well-studied 2D magnetic materials based on $3d$ electrons, experimental studies on rare-earth-based 2D magnetic systems are far fewer, leaving some fundamental questions unanswered. 

Samarium (Sm)-based compounds have long attracted interest in condensed matter physics due to their strong electron correlations and unique quantum phases, including Kondo effects, mixed-valence states, correlated topological phases, self-compensated magnetism, etc \cite{Varma-PRM-1976,Smreview-2018,SmB6PRL2010,SmB6NRP2020,SmAl2Nature1999}. 
For instance, SmB$_{6}$ has been identified as a topological Kondo insulator, where the interactions between $4f$ electrons and conduction bands gives rise to opening of a bulk gap, with topologically nontrivial surface states \cite{SmB6PRL2010,SmB6NRP2020}. On the other hand, SmS exhibits strong valence mixing and Kondo lattice behaviors under high pressure \cite{SmSPRL2015}, highlighting the strong electron correlations in Sm-based compounds.
Another peculiar property of Sm-based compounds is that Sm can exhibit self-compensated magnetism, where the spin and orbital magnetic moments are of similar magnitude and align in opposite directions: SmAl$_{2}$ is a well-known example of such compensated ferromagnetism, which shows zero net moment at certain temperature due to the delicate balance between its spin and orbital contributions \cite{SmAl2Nature1999}. 
These intriguing characteristics make Sm-based compounds an exciting platform for exploring 2D materials.

Despite aforementioned interesting properties, experimental studies on Sm-based compounds have primarily focused on their macroscopic physical properties, with relatively few investigations into their electronic structures, except for the topological {kondo} insulator candidate SmB$_{6}$ \cite{SmB6ARPESreview2014,SmB6NC2013,JiangSmB6NC2013,SmB6PRB2014-MShi}. In particular, there are very few studies on the electronic structure of Sm-based magnetic materials, hindering an in-depth understanding of the electronic origin of the magnetism caused by Sm $4f$ electrons. In this paper, we choose to study a Sm-based magnetic material SmC$_{6}$ \cite{SmC6PRB1994}, which is essentially a 2D material made from Sm intercalation of graphene layers. 
We successfully grow ultrathin SmC$_{6}$ layers on graphene-terminated SiC substrates using molecular beam epitaxy (MBE). $In$ $situ$ measurements from angle-resolved photoemission spectroscopy (ARPES) reveal valence bands of SmC$_{6}$, consistent with density functional theory (DFT) calculations assuming localized $4f$ electrons. 
The valence bands exhibit large band splitting that can be attributed to the ferromagnetic order. The size of splitting is up to 0.16 eV, which is comparable in size to those in Eu- and Gd-based ferromagnets with much larger local moments. 
Interestingly, the surface state also exhibits band splitting with similar magnitude and can be eliminated by overannealing. 
Our work provides valuable insights to understand the electronic origin of magnetism in Sm-based materials. Our study also opens opportunities to study 2D magnetic materials using Sm-based compounds.

\section{Method}
\subsection{Experiment}
\begin{figure}[ht]
\centering
\includegraphics[width=1.0\linewidth]{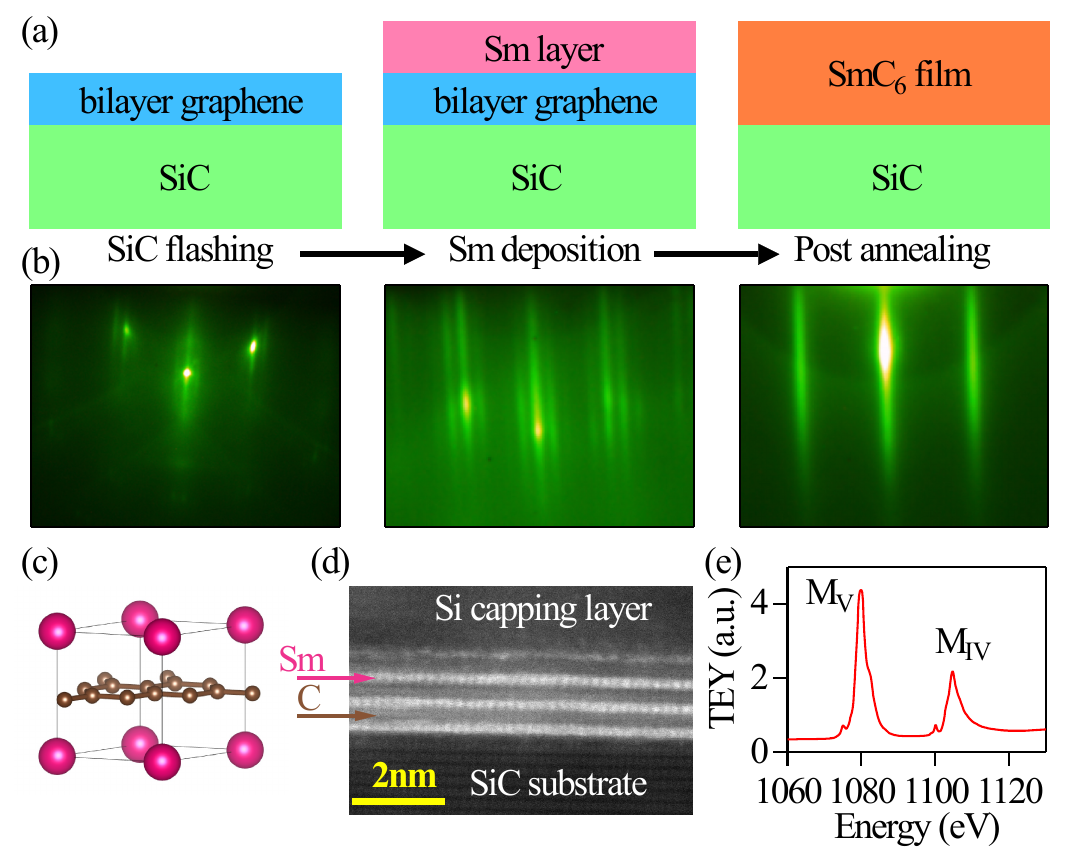}
\caption{Growth and characterization of SmC$_{6}$ thin films. 
(a) Schematic illustration of the growth process of epitaxial SmC$_{6}$ films. 
(b) \textit{In-situ} RHEED patterns corresponding to each stage in (a). 
(c) Crystal structure of bulk SmC$_{6}$ consisting of alternating Sm and graphene layers.  
(d) Cross-sectional HAADF image of SmC$_{6}$ films capped with a Si layer, obtained from the STEM mode. The brighter and weaker atoms at the interface correspond to Sm (pink arrow) and C (brown arrow), respectively.  
(e) XAS spectrum of a SmC$_{6}$ film taken with total electron yield (TEY), highlighting the Sm $M_{V}$ and $M_{IV}$ peaks associated with Sm$^{3+}$. 
\label{Fig1}}
\end{figure}

The growth recipe of epitaxial SmC$_{6}$ films is illustrated schematically in Fig. \ref{Fig1}(a). Prior to film growth, the Si-terminated 6H-SiC(0001) substrates were prepared by direct-current heating up to 1400°C, to produce few-layer epitaxial graphene. Sm layers were then deposited onto the substrates at room temperature using an effusion cell, with a deposition rate of approximately $\sim$1 Å/min as calibrated by a quartz crystal monitor (QCM).
Following the deposition of Sm layers (typically a few nm), the samples were annealed at $\sim$300°C for about 5 mins, to remove excess Sm atoms and facilitate the intercalation of Sm into the graphene layers.
The optimal annealing temperature ($\sim$300°C) for the best SmC$_{6}$ films was determined by $in$ $situ$ measurements from reflection high-energy electron diffraction [RHEED, see Fig. \ref{Fig1}(b)] and ARPES (see below). The samples in Fig. \ref{Fig5} were annealed to $\sim$350°C. 
The base pressure of the MBE chamber was below $1\times10^{-10}$ mbar, which increases to 5-10$\times10^{-10}$ mbar during the deposition of Sm layers.

After growing SmC$_{6}$ films, the samples were transferred under ultrahigh vacuum to the ARPES chamber for electronic structure measurements. The base pressure of the ARPES system was approximately 6$\times10^{-11}$ mbar, increasing to 1.5$\times10^{-10}$ mbar during helium lamp operation. 
Both He I$\alpha$ (21.2 eV) and He II (40.8 eV) photons were utilized in the measurements. 
The typical energy and momentum resolutions were $\sim$12 meV and $\sim$0.01 {\AA}$^{-1}$, respectively.

$Ex$ $situ$ scanning transmission electron microscopy (STEM) measurements using the high-angle annular dark field (HAADF) imaging were performed for SmC$_{6}$ films capped with 10 nm amorphous Si layers, which were deposited in the MBE chamber at room temperature. The Si capping layers were used to protect the SmC$_{6}$ films from oxidation in ambient conditions. A state-of-the-art spherical aberration-corrected transmission electron microscope was used for the STEM measurements, enabling atomic-resolution imaging of the film structure. X-ray absorption spectroscopy (XAS) measurements near the Sm $M$ edge were performed on samples capped with a 5 nm Si layer. The XAS measurements were performed at the BL07U beamline of the Shanghai Synchrotron Radiation Facility (SSRF) \cite{XMCDbeamline}. The XAS experiments were performed at a sample temperature of 10 K.

\subsection{{First-Principles Calculation}}
DFT calculations were performed using the Vienna $ab$ $initio$ Simulation Package (VASP) \cite{VASP1993}, equipped with the projector-augmented wave method and the Perdew-Burke-Ernzerhof exchange-correlation functional \cite{PAW1999,PBE1996}. 
Previous study has shown that bulk SmC$_{6}$ has a hexagonal structure with alternating graphene and Sm layers \cite{SmC6PRB1994}, and the Sm atoms form a triangular lattice. The bulk crystal structure is illustrated in Fig. \ref{Fig1}(c). The lattice parameters were set to be $a = b = 4.314 \AA$, $c = 4.872\AA$, which are consistent with STEM measurements [Fig. \ref{Fig1}(d)]. The energy cutoff for all calculations was fixed at 600 eV. 
Since the Sm $4f$ electrons are very localized, the "Sm\_3" pseudopotential in VASP was employed, which assumes a $4f^5$ configuration for Sm (or Sm$^{3+}$). Therefore, the $4f$ electrons are treated as core electrons and are not included in the DFT calculations, and this corresponds to a nonmagnetic state, namely, data above transition temperature in our experiments.

\section{Results and Discussion}
\subsection{Growth and Characterization of SmC$_{6}$ Films}

The $in$ $situ$ RHEED patterns taken at each stage of film growth are shown in Fig. \ref{Fig1}(b). After room-temperature deposition of Sm layers, the RHEED pattern becomes streaky with a strong $5 \times 5$ reconstruction, consistent with previous studies \cite{Smsurface-PRL1989,Smsurface-PRL2002}. After further annealing at $\sim$300°C, a sharp 1x1 pattern corresponding to the bulk lattice of SmC$_{6}$ can be obtained, indicating formation of high-quality SmC$_{6}$ films. Detailed analysis of the RHEED linecuts shows that the in-plane lattice constant of SmC$_{6}$ films is $4.29 \AA$, very close to the bulk value reported previously. The HAADF-STEM image of a typical SmC$_{6}$ film (capped with Si layers) is shown in Fig \ref{Fig1}(d), which clearly shows a sharp interface between the SiC substrate and SmC$_{6}$ film. The Sm and C layers in the SmC$_{6}$ film can be identified directly in the STEM image, with Sm and C atoms showing up as brighter and weaker dots, respectively. The thickness of the SmC$_{6}$ films is determined to be $\sim$3-4 unit cells. The signal from the top SmC$_{6}$ layer is noticeably weaker than that of the underlying layers. This attenuation may result from spatial inhomogeneity of the SmC$_{6}$ film or the influence from the Si capping layer.

In Sm-based compounds, the valence of Sm is important to determine their physical properties at low temperatures. For divalent Sm ($4f^6$), Hund's rules predict $J$ = 0, resulting in a nonmagnetic ground state. In contrast, trivalent Sm ($4f^5$) has a total angular momentum $J$ = 5/2, yielding a finite effective magnetic moment of $\sim$0.84 $\mu_B$. Therefore, trivalent Sm (Sm$^{3+}$) leads to the formation of local moments, which can give rise to long-range magnetic order at low temperatures via Ruderman-Kittel-Kasuya-Yosida (RKKY) interactions. To investigate the valence of Sm in SmC$_{6}$ films, we performed synchrotron-based XAS measurements at the $M$ edge of Sm [Fig. \ref{Fig1}(e)]. The XAS spectrum clearly demonstrates a dominant contribution from Sm$^{3+}$ \cite{Sm0.3Y0.7S-JAP1984,SmCoarxiv2025}, from both the $M_{V}$ and $M_{IV}$ edges. The trivalent Sm in SmC$_{6}$ films validates our core-$4f$ treatment in DFT calculations and also allows for the development of ferromagnetic order at low temperatures, which we shall discuss in details below.

\subsection{Electronic Structure of SmC$_{6}$ Films and Its Temperature Evolution}
\begin{figure}[ht]
\includegraphics[width=1.0\columnwidth]{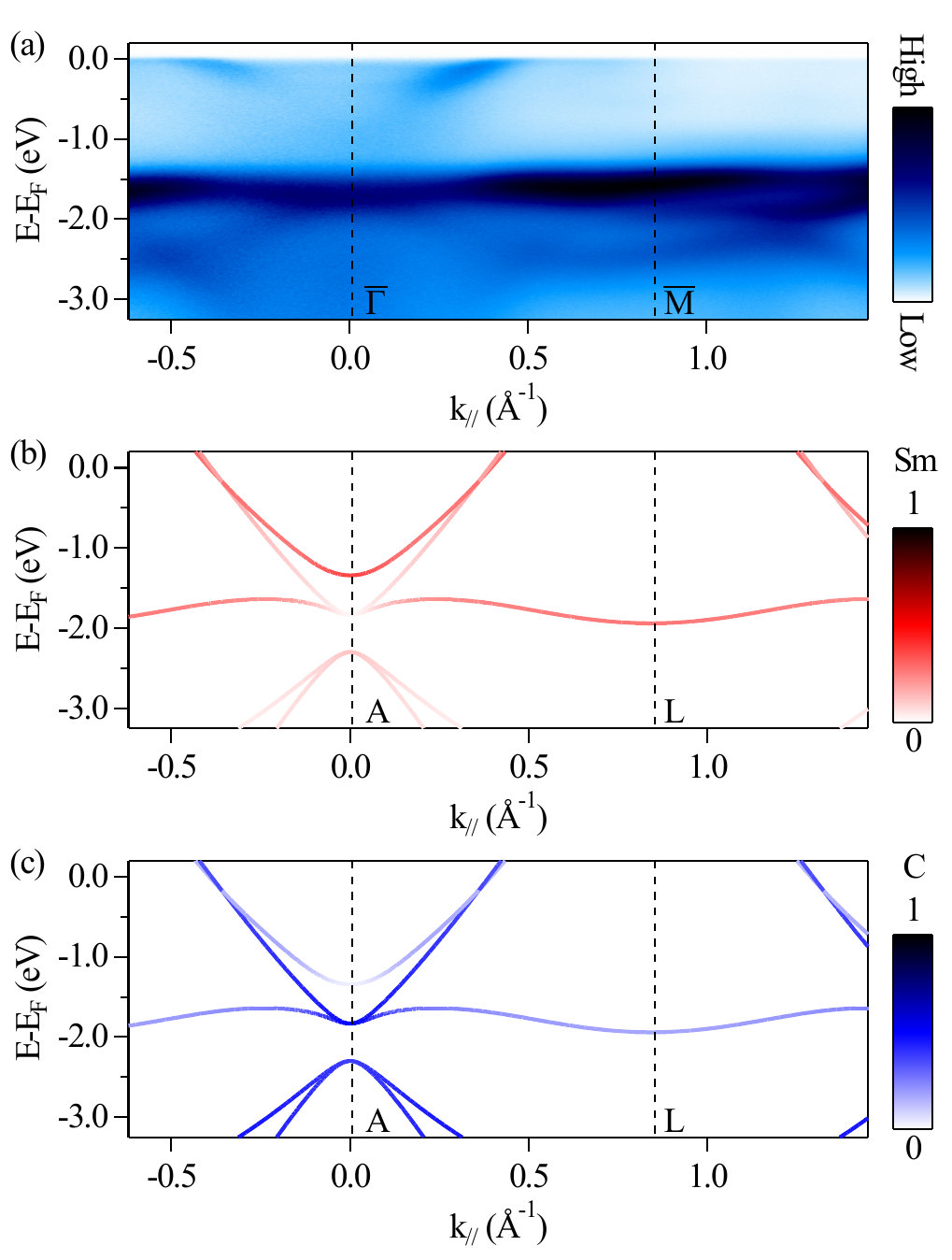}
\centering
\caption{Electronic structure of SmC$_{6}$ films at 125 K and comparison with DFT.
(a) ARPES spectrum of a SmC$_{\rm6}$ film along the $\bar{\Gamma}-\bar{M}$ direction, measured with 21.2 eV photons.  
(b,c) Calculated band structure of SmC$_{6}$ at $k_{z}$=$\pi$/$c$ from DFT, with projected contributions from Sm (b) and C (c), respectively. 
}
\label{Fig2}
\end{figure}

\begin{figure}[ht]
\centering
\includegraphics[width=1.0\columnwidth]{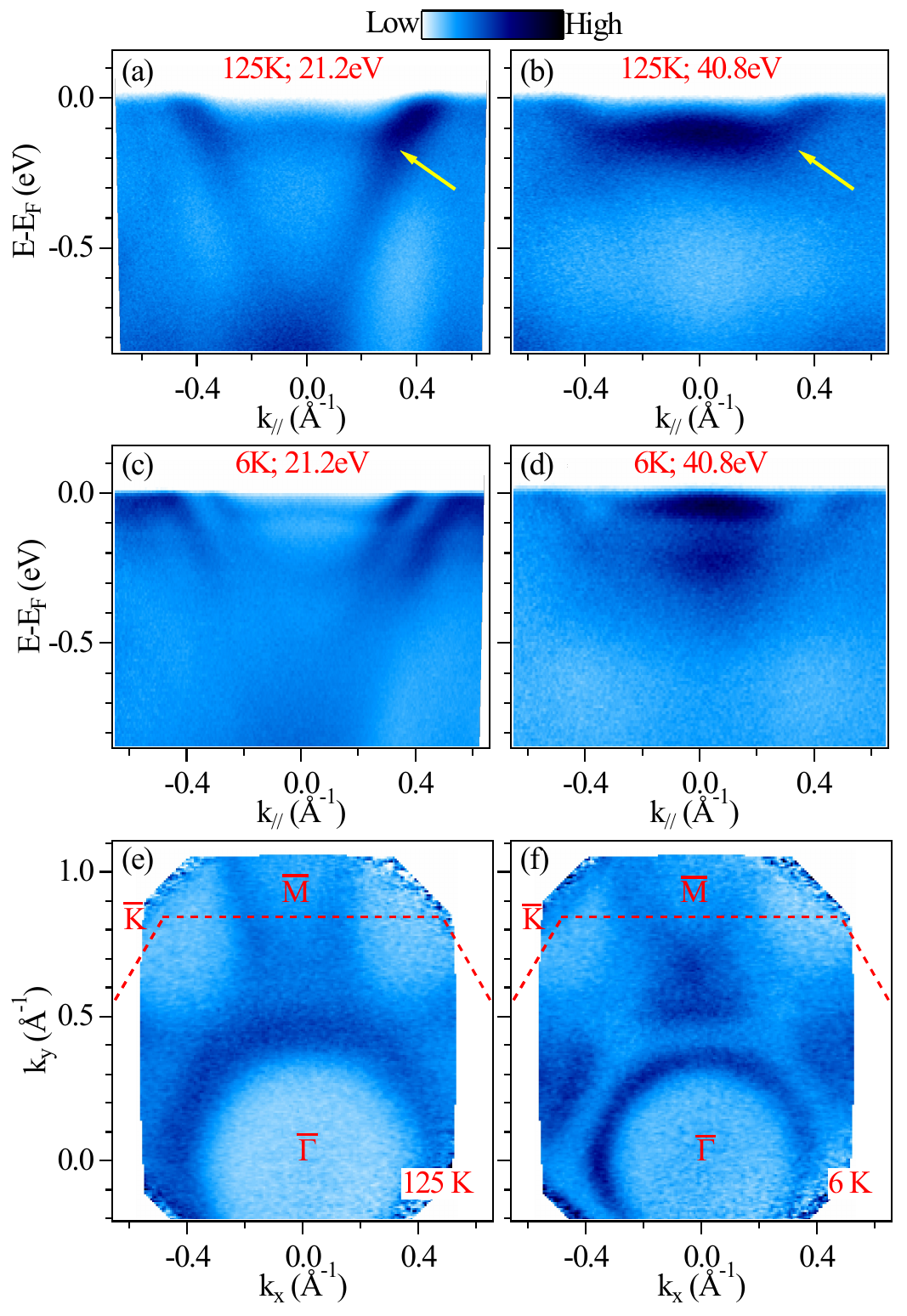}
\caption{Temperature-dependent band structure of SmC$_{6}$ films.  
(a,b) ARPES spectra along $\bar{\Gamma}-\bar{M}$ measured at 125 K using 21.2 eV (a) and 40.8 eV (b) photons, respectively.  
(c,d) ARPES spectra along $\bar{\Gamma}-\bar{M}$ measured at 6 K using 21.2 eV (c) and 40.8 eV (d) photons, respectively.  
(e,f) Fermi surface maps at 125 K (e) and 6 K (f) measured with 21.2 eV photons.
}
\label{Fig3}
\end{figure}

Now we turn to the electronic structure of SmC$_{6}$ films. Figure \ref{Fig2}(a) shows the ARPES spectrum at 125 K taken with 21.2 eV photons, and Fig. \ref{Fig2}(b,c) displays the calculated band structure from DFT for comparison. Here $k_z = \pi/c$ is adopted in the DFT calculation (corresponding to an estimated inner potential of 22.9 eV), as this $k_z$ cut gives the best agreement with the experimental data. Specifically, near the $\bar{\Gamma}$ point, the large electron pocket crossing the Fermi level ($E_F$), the nearly flat band around $-1.6$ eV and the hole-type band below $-2$ eV can be well reproduced by the DFT calculation. 
Note that the flat band near $-1.6$ eV is not derived from Sm $4f$ electrons, as the $4f$ states corresponding to Sm$^{3+}$ typically lie 4 eV below $E_F$ \cite{SmB6ARPESreview2014,SmRh2Si2-PRB2017} - see also our core-level measurements in Fig. \ref{Fig5}(c). 
This is also supported by our atom-resolved band structure as shown in Fig. \ref{Fig2}(b,c): both Sm and C atoms contribute significantly to this band. 
Near $E_F$, large electron pockets centered at $\bar{\Gamma}$ are predicted by DFT and they are indeed observed experimentally (Fig. \ref{Fig2}(a) and Fig. \ref{Fig3}(a)). In general, the experimental electronic structure at 125 K shows good agreement with DFT calculations.

Figure \ref{Fig3}(a) shows an enlarged view of valence bands near $E_F$ at 125 K, obtained from 21.2 eV photons. While a large electron pocket centered at $\bar{\Gamma}$ can be observed, consistent with the DFT calculation in Fig. \ref{Fig2}(b), there is an additional weak hole band with its band top at $\sim$$-0.1$ eV, which cannot be explained by the DFT calculation. As we shall show later, this hole band can be destroyed by overannealing the sample and forming a different reconstructed surface. Therefore, it can be attributed to surface state that is not included in the DFT calculation. When the large electron band crosses the hole band at $k_{//}$ $\sim$ $\pm$ 0.3 Å$^{-1}$, kinks develop in its dispersion, as indicated by a yellow arrow. 
Fig. \ref{Fig3}(b) shows the same in-plane momentum cut taken with 40.8 eV photons, where the surface hole band at $\sim$$-0.1$ eV becomes much stronger, but the bulk-derived electron band becomes obviously weaker. 
Note that the ARPES data taken with 40.8 eV photons contain strong contributions from the surface state, which overwhelm the bulk states and make it difficult to make a reliable comparison with the bulk DFT calculations. 
Nevertheless, the unusual dispersion and intensity distribution of the electron band shown in Fig. \ref{Fig3}(a,b) indicate fine interactions between the electron and hole bands.

Lowering the temperature from 125 K to 6 K, both the electron band and hole band split into two bands, with an energy separation up to 0.16 eV, as shown in Fig. \ref{Fig3}(c) and (d). Note that the splittings are of similar size in both 21.2 eV and 40.8 eV spectra, confirming that the splittings are intrinsic. The hybridization between the electron and hole bands with large splittings leads to additional band bendings and gap openings at low temperature, which can be best seen in the 21.2 eV data [Fig. \ref{Fig3}(c)]. For energies away from $E_F$, the bands become more diffuse, probably due to the shorter quasiparticle lifetimes at higher binding energies. For the electronic bands below -1 eV, no obvious change is observed from 125 K down to 6 K. Fig. \ref{Fig3}(e) and (f) display the Fermi surface maps at 125 K and 6 K, respectively. At 125 K, a large pocket centered at $\bar{\Gamma}$ can be observed. It splits into two pockets at 6 K, which consist of a small circular pocket and a large hexagonal pocket that further hybridizes with the pockets centered at $\bar{K}$ and forms fine structures along $\bar{\Gamma}-\bar{M}$ [Fig. \ref{Fig3}(f)]. The Fermi surface evolution is consistent with the band dispersion shown in Fig. \ref{Fig3}(c), confirming the Zeeman-type band splitting.

\begin{figure*}[ht]
\includegraphics[width=1.0\linewidth]{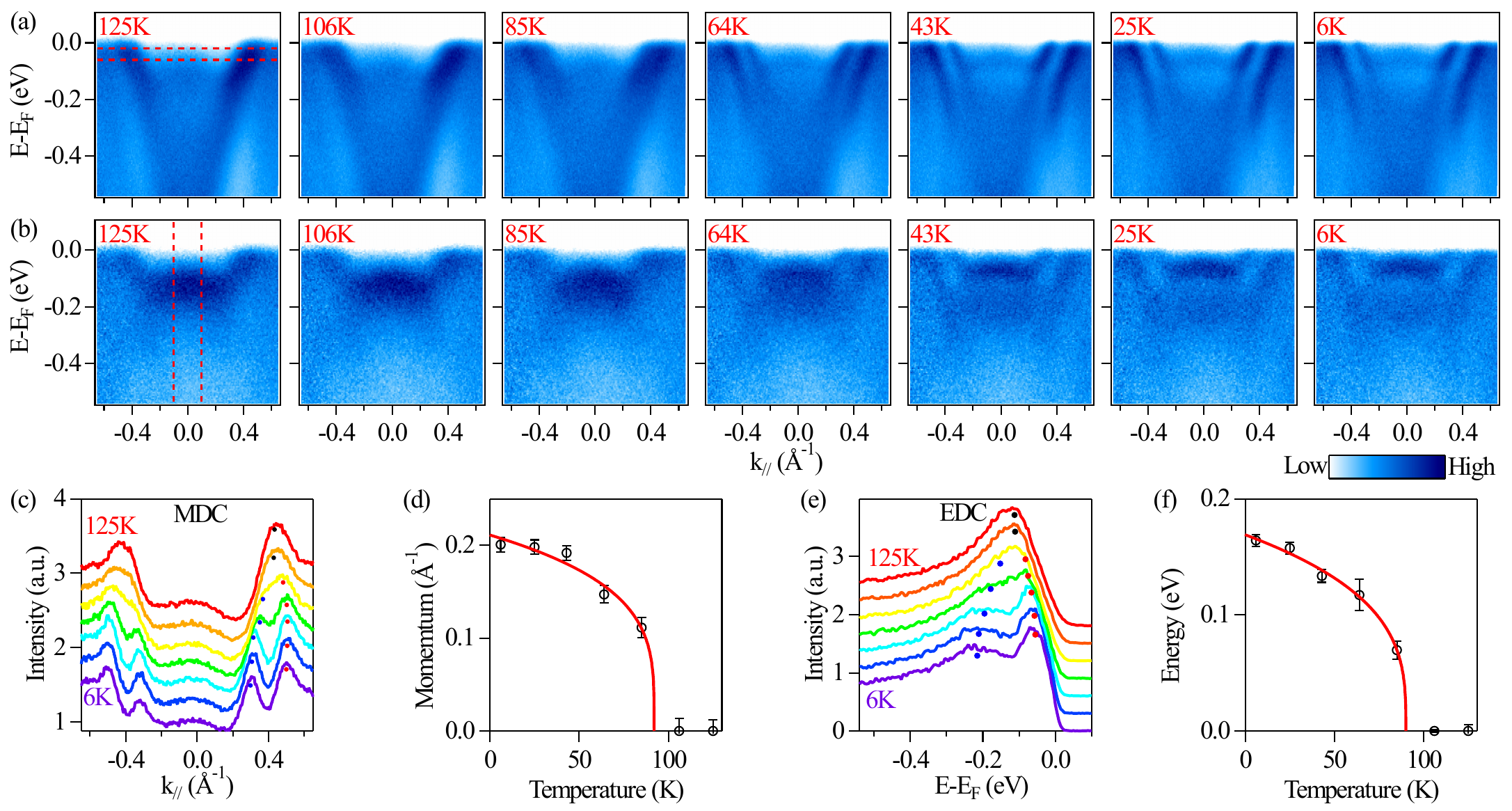}
\centering
\caption{Detailed temperature evolution of valence bands near $E_F$ and analysis of band splittings. 
(a,b) Detailed temperature evolution of the ARPES spectra near $E_F$ measured with 21.2 eV (a) and 40.8 eV (b) photons. The data were obtained from the same sample during two separate cooling cycles.
(c) Temperature-dependent MDCs at $E$ = $-0.04$ eV extracted from (a), integrated over the energy range indicated by the red dashed box in (a). The fitted peak positions are labeled as blue and red dots (below 100K) or black dots (above 100K) in MDCs. 
(d) Extracted momentum separation from (c) as a function of temperature and power-law fitting of data. 
(e) Temperature-dependent EDCs near $\bar{\Gamma}$ extracted from (b), integrated over the momentum range indicated by the red dashed box in (b). The fitted peak positions are labeled as blue and red dots (below 100K) or black dots (above 100K) in EDCs. 
(f) Extracted energy separation from (e) as a function of temperature and power-law fitting of data. 
Error bars are included both in (d) and (f). 
}
\label{Fig4}
\end{figure*}

Detailed temperature evolution of the ARPES spectra taken with 21.2 eV and 40.8 eV photons is summarized in Fig. \ref{Fig4}(a,b). To analyze the temperature evolution of the electron and hole bands, we extract temperature-dependent momentum distribution curves (MDCs) from the 21.2 eV data and energy distribution curves (EDCs) from the 40.8 eV data, by integrating the red dashed boxes in Fig. \ref{Fig4}(a,b). The results, shown in Fig. \ref{Fig4}(c,e), illustrate the temperature-dependent splitting of the electron and hole bands, respectively. For quantitative analysis, each curve at one temperature was fitted using two Lorentzian functions with a quadratic polynomial background, unless at temperatures above 100 K, where no discernible splitting can be identified. The fittings yield the peak positions and their momentum or energy splittings as a function of temperature, summarized in Fig. \ref{Fig4}(d,f). The results show that band splittings already occur at $\sim$85 K and the momentum/energy splittings exhibit a obvious temperature dependence. Detailed fittings using power-law relation (red line in Fig. \ref{Fig4}(d,f)) agree well with experimental results, demonstrating a mean-field-like behavior, with estimated onset temperature of $\sim$90 K both for the electron and hole bands.

Considering the trivalent nature of Sm in SmC$_{6}$ films from the XAS measurements, the most natural explanation for the observed band splittings is the ferromagnetic exchange splittings arising from Sm$^{3+}$ moments. Specifically, the Sm local moments develop long-range ferromagnetic order below 90 K through the RKKY interactions mediated by the conduction electrons, leading to Zeeman-type spin splitting of conduction bands. The mean-field-like temperature evolution of splittings provides additional support for the (second-order) ferromagnetic transition. We mention that Sm-based ferromagnetic compounds with comparable ordering temperatures have been well documented in the literatures, e.g., SmAl$_{2}$ with a T$_{C}$ of $\sim$120 K \cite{SmAl2Nature1999}, SmCd with a T$_{C}$ of $\sim$195 K and SmZn with a T$_{C}$ of $\sim$128 K \cite{PhysRevB.59.11445}. The gradual temperature evolution of the splittings (Fig. \ref{Fig4}) and the Zeeman-type isotropic splittings in momentum space (Fig. \ref{Fig3}) also rule out other alternative explanations for the observed band splittings, such as quantum well states in ultrathin films \cite{ChiangPhotoemission}, charge/spin density waves, satellites from electron-boson coupling \cite{Lee2014} and altermagnetic splittings in compensated antiferromagnets \cite{PhysRevX.12.040501,BaiAFM2024,Song2025}. Therefore, we believe that ferromagnetic splitting is the most natural and plausible explanation for the observed band splitting in SmC$_{6}$ films.

It is interesting to see that, despite the small magnetic moment of Sm$^{3+}$ ions (0.84 $\mu_B$), the observed magnetic splitting in SmC$_{6}$ films can be up to 0.16 eV (Fig. \ref{Fig4}(d,f)), comparable in magnitude to that of Eu- and Gd-based systems \cite{EuRh2Si2-NC2014,GdRh2Si2-SR2016,EuCd2P2-PRB2023}, where the Eu$^{2+}$ or Gd$^{3+}$ local moments ($4f^7$) are typically one order of magnitude larger. Since the size of conduction band splitting due to ferromagnetically ordered $f$-electron local moments should be proportional to the product of the magnitude of local moment and the exchange coupling $J$ between $4f$ and conduction electrons \cite{REMagnetism}, the surprisingly large ferromagnetic splitting observed in SmC$_{6}$ implies large exchange coupling between Sm $4f$ electrons and conduction bands. It would be very interesting in the future to understand the microscopic origin of such a large exchange coupling in SmC$_{6}$, and whether such a large coupling is also present in other Sm-based compounds. 

We mention that there has been no report of magnetic band splittings in Sm-based compounds from electron spectroscopic measurements, to the best of our knowledge. One possible reason is that Sm-based compounds are often prone to valence instability or fluctuations, which can lead to nonmagnetic ground state or magnetically ordered state with low ordering temperatures. Here in SmC$_{6}$, Sm is predominantly trivalent and therefore the magnetic exchange interactions can be strong. In addition, our synthesis of high-quality epitaxial films by MBE and $in$ $situ$ ARPES measurements allows for direct measurements of the magnetic band splittings.

\subsection{Persistent Magnetic Splitting in overannealed SmC$_{6}$ Films}

\begin{figure}[ht]
\centering
\includegraphics[width=1.0\columnwidth]{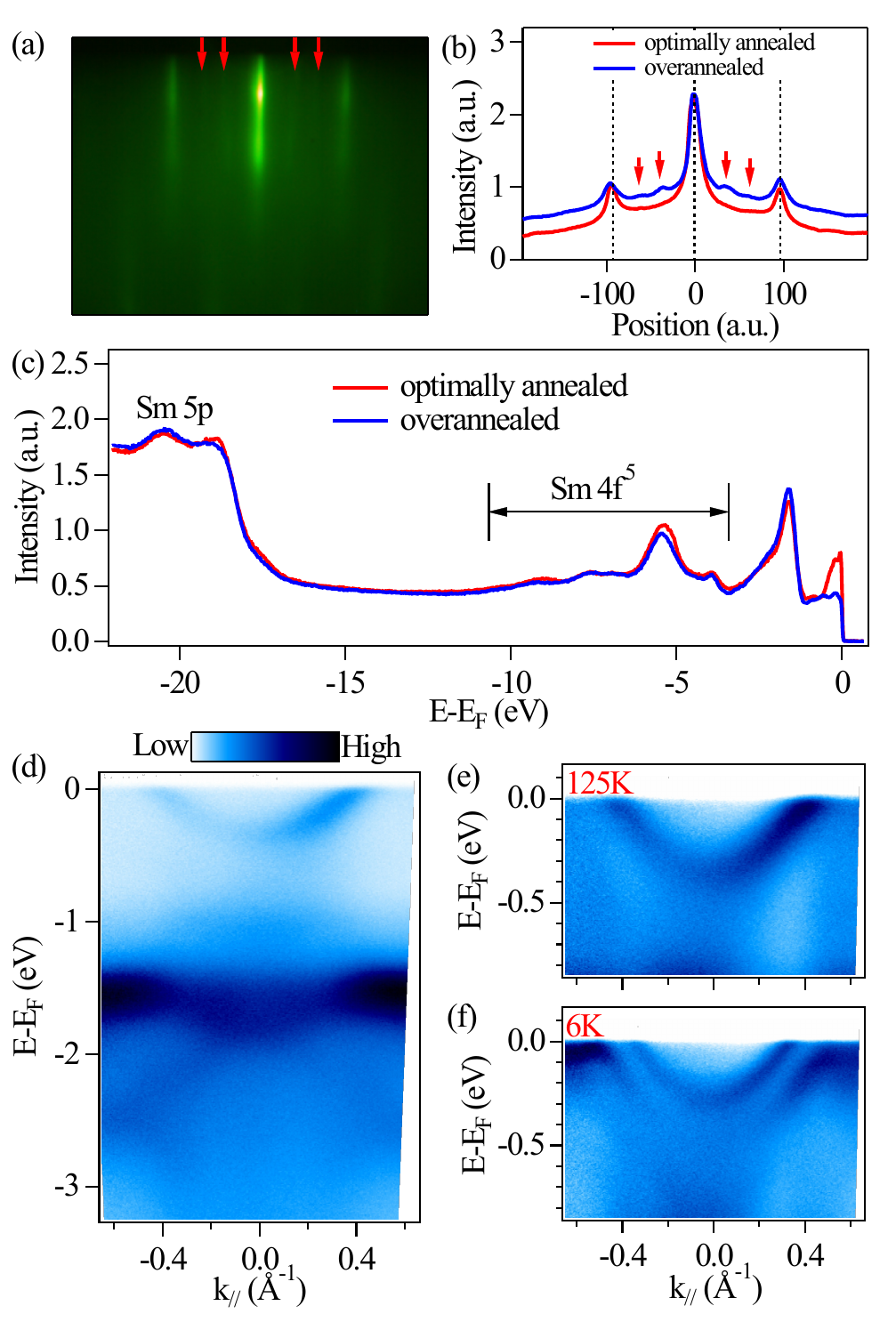}
\caption{Electronic structure of overannealed SmC$_{6}$ films. 
(a) RHEED pattern of an overannealed SmC$_{6}$ film, showing weak superstructure peaks indicated by red arrows. 
(b) Comparison of RHEED line cuts for optimally annealed and overannealed SmC$_{6}$ films. Weak superstructure peaks for the overannealed film are indicated by red arrows.  
(c) Comparison of angle-integrated EDCs for optimally annealed and overannealed SmC$_{6}$ films, taken with 40.8 eV photons. Sm $5p$ and $4f$ core levels are labeled accordingly. 
(d) Valence band of an overannealed film taken with 21.2 eV photons.  
(e,f) Enlarged view of band dispersion for an overannealed film near $E_F$ measured at 125 K (d) and 6 K (e), using 21.2 eV photons.  
}
\label{Fig5}
\end{figure}

Annealing the SmC$_{6}$ films at higher temperatures ($\sim$350°C) leads to weakening of the RHEED patterns and broadening of the ARPES spectra. Nevertheless, these overannealed samples provide useful reference for disentangling the bulk and surface states, as well as testing the robustness of magnetic band splittings. Figure \ref{Fig5}(a) shows the RHEED pattern for an overannealed sample, which reveals weak superstructure peaks (indicated by red arrows) compared to the unreconstructed surface of optimally annealed samples. Figure \ref{Fig5}(b) summarizes the RHEED line cuts of optimally annealed and overannealed samples. The results show that while the in-plane lattice constant remains unchanged with further annealing, two additional superstructure peaks emerge along the $\bar{\Gamma}-\bar{K}$ direction for the overannealed sample. Since no extra superstructure peak is observed along the $\bar{\Gamma}-\bar{M}$ direction, the surface reconstruction corresponds to a $\sqrt{3} \times \sqrt{3}$ $R$30° modulation in real space. 

Overannealing induces surface disorders or creates vacancies on the surface of the SmC$_{6}$ films, but it does not completely change the surface termination or internal structures. This can be verified by the core level scans shown in Fig. \ref{Fig5}(c), where the intensities of the Sm $5p$ states at $\sim$$-20$ eV and the Sm $4f$ states between $-4$ eV and $-12$ eV are not affected much by overannealing, indicating similar surface termination. At the same time, the overall valence bands of the overannealed sample, shown in Fig. \ref{Fig5}(d), show some similarity to those of optimally annealed samples, e.g., the weakly dispersive band near $-1.6$ eV and the electron band centered at $\bar{\Gamma}$ below $E_F$. However, close to $E_F$, some obvious difference can be found after overannealing (Fig. \ref{Fig5}(e)): the hole band originally at $-0.1$ eV near $\bar{\Gamma}$ disappears, and correspondingly the kinks in the electron band are gone, leading to a parabolic electron band at 125 K, as shown in Fig. \ref{Fig5}(e). The disappearance of the hole band in overannealed samples indicate that this band is indeed a surface state that can be destroyed by surface reconstruction. Upon lowering to 6 K, the parabolic electron band splits into two bands, with an energy separation up to 0.16 eV, very similar to that in optimally annealed samples. This demonstrates the robust ferromagnetic splitting in overannealed SmC$_{6}$ films, indicating that the ferromagnetic transition is resilient against surface reconstruction.

\section{conclusion}
In summary, we have successfully grown high-quality SmC$_{6}$ thin films using MBE and investigated their electronic structures by $in$ $situ$ ARPES measurements. 
We observed large band splitting up to 0.16 eV and persisting up to 90 K, implying ferromagnetism in SmC$_{6}$ thin films. 
Despite the small magnetic moment associated with Sm$^{3+}$, the ferromagnetic-like splitting in SmC$_{6}$ films is comparable to those observed in large-moment Eu and Gd-based systems, which indicates a strong exchange coupling $J$ between the Sm $4f$ moments and conduction bands. Interestingly, the ferromagnetic order can lead to Zeeman-type splitting of surface state, with similar size as that of bulk bands. For overannealed samples with surface reconstruction, the surface state disappears, but the ferromagnetic-like band splitting persists for the bulk bands. This suggests a resilient ferromagnet order in SmC$_{6}$ thin films, offering another platform to study two-dimensional magnetic materials. The insight into the electronic structure also enriches our understanding of the interaction between localized $4f$ moments and conduction electrons in Sm-based magnetic materials.

\section*{acknowledgments}
This work is supported by the National Key R$\&$D Program of China (Grant No. 2022YFA1402200, No. 2023YFA1406303), 
the Key R$\&$D Program of Zhejiang Province, China (2021C01002), 
the National Natural Science Foundation of China (No. 12174331, No. 12350710785) 
and the State Key project of Zhejiang Province (No. LZ22A040007), 
and the Zhejiang Provincial Natural Science Foundation of China (Grant No. LGG22E020006).
We also acknowledges the reform research fund of Teaching Guidance Committee for Physics Majors from the Ministry of Education (Grant No. 2024PRO33).
We thank Prof. Denis V. Vyalikh, Prof. Zhentao Wang, Prof. Yi Yin, Prof. Yuanfeng Xu, Dr. Ge Ye, Mr. Jianzhou Bian and Mr. Yonghao Liu for discussions and help on experiments. We also thank the staff from BL07U of SSRF for assistance of XAS data collection.

\section*{Data availability}
The data that support the findings of this article are openly available \cite{SmC6data}.

\bibliography{SmC6ref}
\end{document}